# Topological Phase Transition in a Quasi Two-Dimensional Coulomb Gas

A. Gama Goicochea[*]

División de Ingeniería Química y Bioquímica, Tecnológico de Estudios Superiores de Ecatepec, Ecatepec de Morelos, Estado de México 55210, Mexico

and

Z. Nussinov

Department of Physics, Washington University, St. Louis, Missouri 63130, USA

## Abstract

A system with equal number of positive and negative charges confined in a box with a small but finite thickness is modeled as a function of temperature using mesoscale numerical simulations, for various values of the charges. The Coulomb interaction is used in its three-dimensional form, $U(r) \sim 1/r$. A topological phase transition is found in this quasi 2d system. The translational order parameter, spatial correlation function, specific heat, and electric current show qualitatively different trends below and above a critical temperature. We find that a 2d logarithmic Coulomb interaction is not essential for the appearance of this transition. This work suggests new experimental tests of our predictions, as well as novel theoretical approaches to probe quasi 2d topological phase transitions.

---

[*] Corresponding author. Electronic mail: agama@alumni.stanford.edu

# I INTRODUCTION

The conventional 2d Coulomb gas consists of logarithmically interacting charges [1–3]. Kosterlitz and Thouless (KT) showed that such a system undergoes a topological phase transition as a function of temperature [1]. As the system is heated, a transition appears between its low temperature dielectric phase and a conducting high temperature phase [1]. The transition temperature is $T_{KT}^* = q^{*2}/4$, where $q^*$ is the charge, in the low-density limit [4]. Asterisked symbols represent adimensional quantities. The properties of the 2d Coulomb gas have been studied theoretically [1–8] and with numerical simulations [9–17]. At temperatures below $T_{KT}^*$, correlations drop algebraically with distance, with a temperature dependent exponent. By contrast, at temperatures above $T_{KT}^*$, the correlations decay exponentially [1]. The KT transition has long been viewed as the quintessential "topological phase transition" that occurs despite the absence of a well–defined local order parameter (forbidden by the Mermin–Wagner–Coleman theorem [18]). Here, at finite temperature, no continuous symmetry breaking occurs. However, at temperatures $T^* < T_{KT}^*$, the correlations can become very long ranged displaying the said algebraic decay with distance. Topological transitions also appear in gauge theories that model fundamental interactions. By Elitzur's theorem [19] no local order parameter can appear since gauge symmetry may not be broken. The latter transitions in gauge theories are typically discontinuous (first order) or conventional continuous (i.e., second order) transitions [20, 21]. Both KT transitions and phase transitions in gauge theories occur from a confined phase (of particles or topological defects, such as vortices or other excitations) to a deconfined phase. The KT transition is unique in various ways. Perhaps most strikingly, it is neither a conventional first nor second order transition but is rather an endpoint of a continuous line of critical points. Along this critical line, at $T^* < T_{KT}^*$, the system exhibits the above noted KT algebraic decay of correlations. There are other 2d systems that undergo KT type transitions, of which we note only a few. These include point charges [22], a solid with dislocations [23] and perhaps most well–known, the *XY* spin model (including recent experimental observation in magnetic crystalline materials [24]) and superfluid films [25]. It also appears in thin superconducting films where the interaction between vortices is logarithmic at short range [26]. KT transitions are ubiquitous to $p > 5$ state 2d classical clock models [27] and related 1d clock models [28] and feature prominently in quantum spin chains and 2d quantum dimer models [29, 30]. Typically, in one form or another, all equilibrated systems that display a KT transition can be mapped onto an effective 2d plasma of interacting charges. However, with the exception of gauge theories, all of the above noted studies were carried out for strictly 2d systems. Indeed, excusing Renormalization Group considerations for finite thickness systems and the investigation of truly higher dimensional driven non–equilibrium *XY* models featuring KT–type transitions [31], the detailed nature of the transitions in equilibrated quasi-2d systems (in particular, those in hard/soft sphere charged fluids that form the focus of our work [5, 32]) has long remained largely unexplored.

**II MODEL**

Here we study a low density, quasi – 2d Coulomb gas of spheres confined to move in a box whose thickness is small but finite, by means of numerical simulations. The specific query that motivated our study is that of determining how the topological transition of pristine 2d systems evolves as these acquire a thickness in a transverse direction [33]. That is, what transpires when instead of having charges (either exact or effective, e.g., vortices or dislocations) interacting via 2d, logarithmic–type interactions, the system is a thin layer, and the particles are 3d spheres? In investigating this question, we found that various features persist. Our results may afford a more direct comparison with experiments–which are not truly 2d [34–36]. The large-distance electrostatic interaction that we focus on is, accordingly, modeled as the typical 3d electrostatic interactions, $U(r) \sim 1/r$ where $r$ is the relative distance separating the charges, instead of $U(r) \sim ln(1/r)$ for strictly 2d systems. These systems are neutral, with an equal number of positively and negatively charged particles. For the non–electrostatic interactions, we use the dissipative particle dynamics (DPD) model [37, 38], which is driven by conservative, dissipative, and random forces. The latter forces are short ranged, repulsive and pairwise additive. The conservative force is given by $\vec{F}_{ij}^C(r_{ij}^*) = a_{ij}^*(1 - r_{ij}^*/r_C^*)\Theta(r_C^* - r_{ij}^*)\hat{r}_{ij}$, where $a_{ij}^*$ is the intensity of the force, which is chosen here as $a_{ij}^* = 78.3$ in dimensionless units [39]. Although this type of force allows for some overlap of the particles, our choice of $a_{ij}^*$ reduces it [40]. The length $r_C^* = 1$ is a cutoff radius and $\Theta(x)$ is the Heaviside step function. The dissipative and random forces are, respectively, given by $\vec{F}_{ij}^D(r_{ij}^*) = -\gamma(1 - r_{ij}^*/r_C^*)^2[\hat{r}_{ij}.\vec{v}_{ij}^*]\Theta(r_C^* - r_{ij}^*)\hat{r}_{ij}$ and $\vec{F}_{ij}^R(r_{ij}^*) = \sigma(1 - r_{ij}^*/r_C^*)\Theta(r_C^* - r_{ij}^*)\xi_{ij}\hat{r}_{ij}$, [38]. Here, $\vec{v}_{ij}^* = \vec{v}_i^* - \vec{v}_j^*$ is the relative velocity between particles $i$ and $j$. Due to the fluctuation–dissipation theorem, $\sigma^2/2\gamma = k_BT$, where $k_B$ is Boltzmann's constant and $T$ is the absolute temperature [38]. The amplitudes $\xi_{ij}$ are randomly drawn from a uniform distribution between 0 and 1. The DPD model is chosen because its simple force laws make it possible to use time steps that are up to three orders of magnitude larger than those used in conventional molecular dynamics. Additionally, the coupling of the DPD dissipative and random forces creates a robust, built-in thermostat which is more stable than those used in atomistic molecular dynamics [44]. To confine the charges so that they move on a thin slit, an effective wall force is applied at both ends of the simulation box along the $z$–direction, given by $\vec{F}_{wall}(z^*) = a_w^*(1 - z^*/z_C^*)\Theta(z_C^* - z^*)\hat{z}$ [41]. The factor $a_w^*$ sets the maximal size of this wall strength that has $z_C^*$ as its cutoff distance. The role of this featureless wall force is only to confine the motion of the charges on a slit. The charges are modeled as charge distributions centered at the center of mass of each particle, given by $\rho_{q^*}^*(r^*) = (q^*/\pi\lambda^{*3}) \exp(-2r^*/\lambda^*)$ [42, 43]. Here, $\lambda^*$ is the decay length of the charge distribution and $q^*$ is the total charge carried by the particle. It has been shown [43] that for $r^* \leq r_C^*$ (the particles' radius), the electrostatic force between these charge distributions tends to zero as $r \to 0$. For $r^* > r_C^*$ the electrostatic potential assumes its usual form, $U(r) \sim 1/r$, and its long-range contribution is calculated using the Ewald sums [44]. Using this charge model with the conservative DPD force, the collapse of particles with opposite charge is avoided, and no singularity occurs in the Coulomb interaction at the shortest distances. The number density, defined as the total number of charged particles $N$ divided by the volume of the simulation box $V$, is in all cases equal to $\rho^* = 0.03$. The thickness of the simulation box is fixed at $L_z^* = 1$ while its square transversal area depends on the total number of charged particles, which goes from $N = 200$, up to $3 \times 10^4$. The simulations are run for up to $4 \times 10^3$ blocks of $10^4 \delta t^*$ each, with the time step set at $\delta t^* = 0.01$. Periodic boundary conditions are applied along the *x*- and *y*- axes but not along the *z*-direction since the walls are impenetrable. Full details about the force field and simulations are provided in the Appendix, where additional results can also be found. Further information about the method, algorithm and other details have been published elsewhere [40, 41, 43].

**III RESULTS AND DISCUSSION**

To track the phase transition, we calculate the translational order parameter (TOP) of the particles, defined as follows [44]:

$$\Psi_T = \frac{1}{N} \langle \left| \sum_{j=1}^{N} e^{i\vec{K}\cdot\vec{r}^*_j} \right| \rangle, \tag{1}$$

where $N$ is the total number of particles, and $\vec{K}$ is the first shell reciprocal lattice vector. The angular brackets denote an average over time. Figure 1(a) shows the dependence of TOP on temperature for increasing values of the charge. We find a transition from an ordered low temperature state to a disordered high temperature state, as shown in Fig. 1(a). The transition temperature $T_c^*$ associated to each charge $|q^*|$ is found to monotonically increase with the magnitude of the charge. Interestingly, *on scaling the temperature by* $T_c^* = q^{*2}/4r_C^*$, *all of the TOP curves collapse into a single universal function,* (Fig. 1(a)). The TOP is approximately constant below $T_c^*$, where the charges are all paired, and it decays sharply at temperatures above $T_c^*$. Although the magnitude of the TOP depends on the size of the system, becoming smaller as the size increases, $T_c^*$ does not scale with the system size (see Fig. 1(b), for $|q^*| = 7$). At the onset of the transition, the cluster formed by the grouping of dipoles begins to dissociate, fragmenting into smaller groups of dipoles [10, 11]. This splintering causes the TOP to decrease rapidly with a small rise of the temperature, reaching a plateau. At the highest temperatures, some charges still appear in isolated dipoles. The TOP is small but remains non-zero at those temperatures. For sufficiently large systems, several clusters of dipolar pairs are formed, giving rise to a phase with algebraic order only. When extrapolated to the thermodynamic limit, our data in Fig. 1(b) suggest that the TOP tends to zero, as expected for a topological phase transition having no local order parameter. For a fixed value of the single particle charge, so long as the system remains globally neutral and confined along the *z*–direction, the transition between a state where all charges are bound to one where they are unbound occurs at the same temperature.

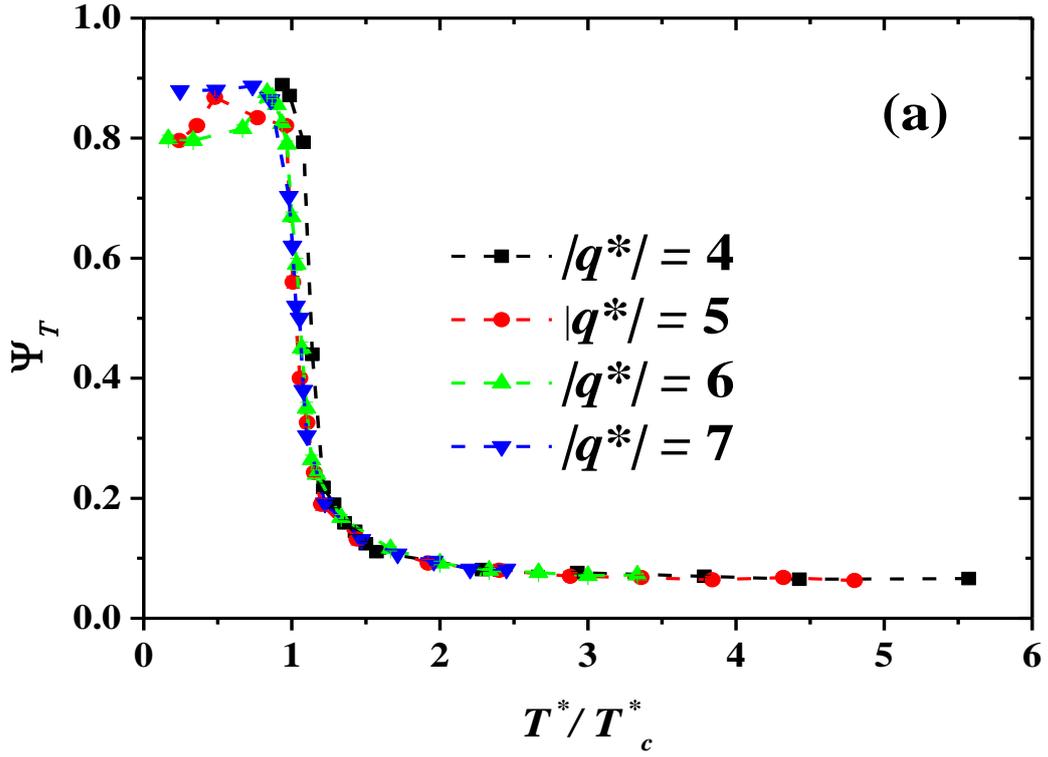

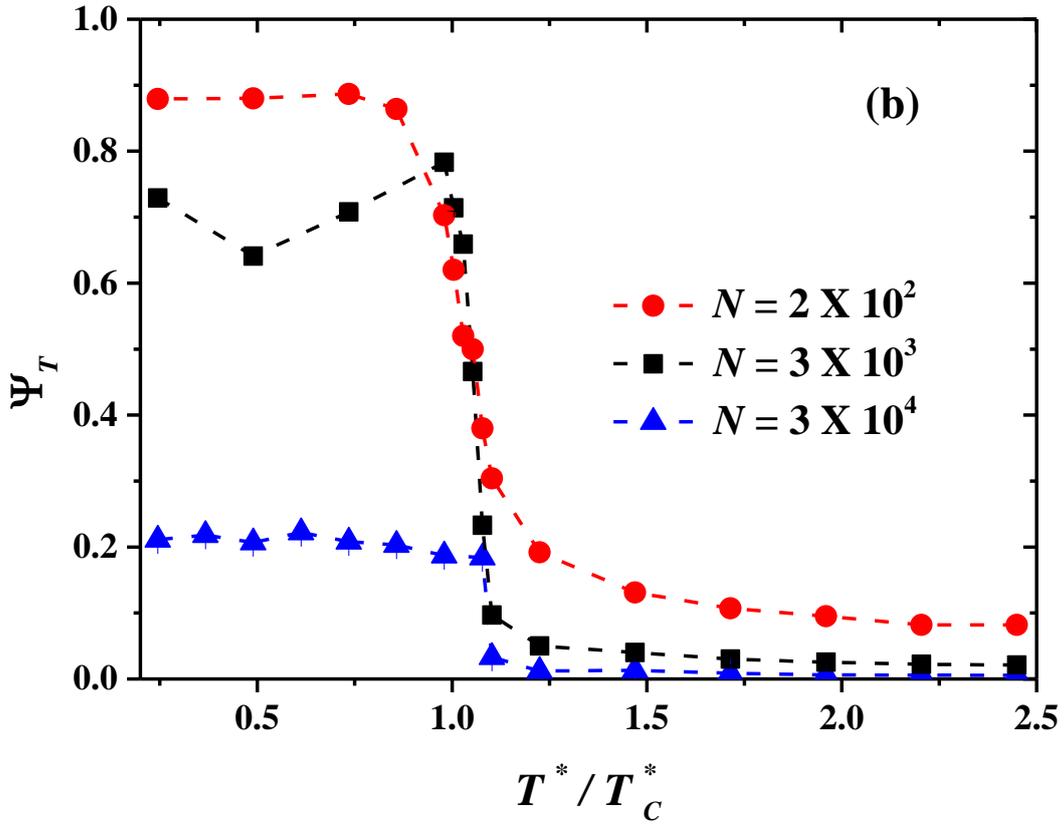

**Figure 1.** (a) The translational order parameter $\Psi_T$ as a function of the scaled temperature, $T^*/T_C^*$ for four values of the charge on the particles. In all cases, $\rho^* = 0.03$, $N = 200$. (b) $\Psi_T$ as a function of the scaled temperature at fixed number density $\rho^* = 0.03$ and charge $|q^*| = 7$ as the total number of charged particles $N$ is increased; error bars are smaller than the symbol's size. The dashed lines are guides for the eye.

The systems modeled in this work belong to the low-density gas regime with long range $1/r$ interactions. As Alastuey and coworkers have argued [16], the transition temperature is a function of the density of the charges through the dielectric function, $\epsilon$, i.e., $T* = q2/4\epsilon$. For a low-density gas, $\epsilon = 1$ and the transition temperature between a low-temperature dielectric phase and a higher temperature conducting one is found to be equal to the KT prediction. For 2d disks with logarithmic electrostatic interactions and hard-disk (contact) non-electrostatic interactions, Orkoulas and Panagiotopoulos [10] find that the transition temperature is density-dependent. We indeed remark that in systems having contact interactions, the system density and thus the local density of contacts may, rather naturally, play a prominent role in determining the transition temperature. At low densities, the critical temperature of [10] tends to that of the KT transition. Using soft DPD interactions and charge distributions between spheres we find that $T_C^*$ is given by the KT prediction for low but finite density.

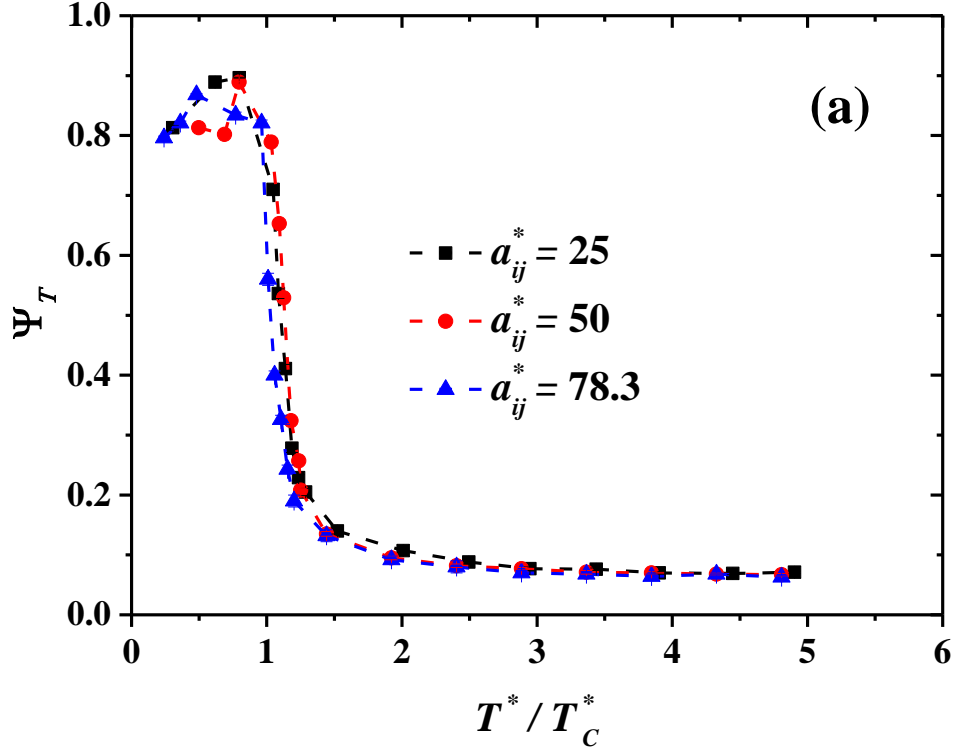

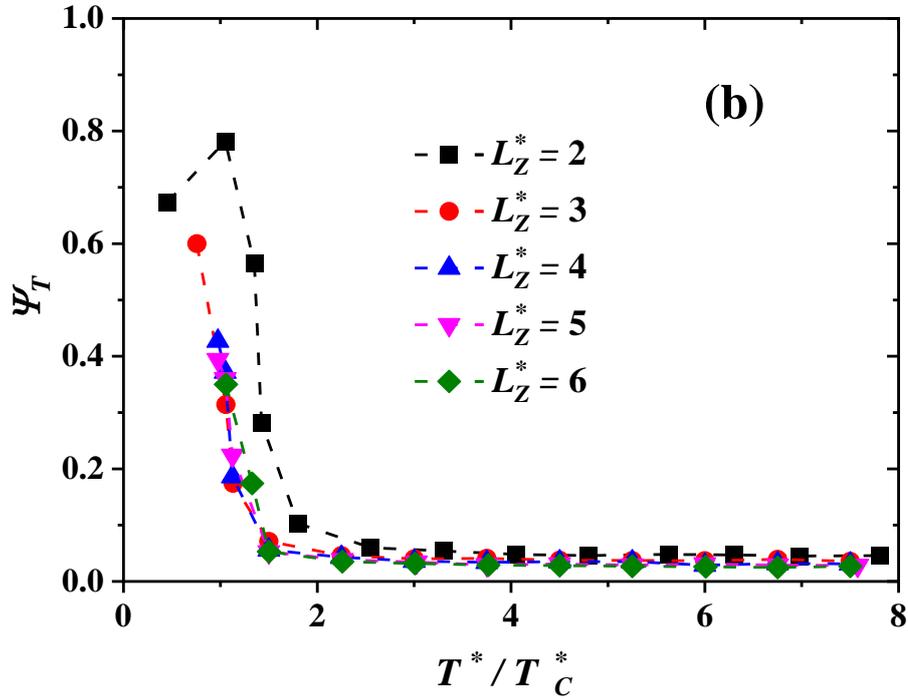

**Figure 2.** (a) Translational order parameter TOP, $\Psi_T$, see eq. (1), *vs* temperature, normalized by $T_c^*$, for increasing intensity of the conservative DPD force constant $a_{ij}^*$ (see eq. (A2a) in the Appendix) with a total of $N = 200$ particles with charge $|q^*| = 5$, in a box with thickness $L_Z^* = 1$. (b) Influence of increasing the thickness of the simulation box along the $z$ – axis, $L_Z^*$, on $\Psi_T$, as a function of the reduced temperature, $T^*/T_C^*$, for $|q^*| =$

4. The transverse area of the box is equal to $L_X^* \times L_Y^* = 80 \times 80 r_C^2$. In all cases, the number density is $\rho^* = 0.03$. Dashed lines are only guides for the eye.

Figure 2(a) shows how the TOP changes as the intensity of the non-bonding, non-electrostatic conservative DPD interaction, $a_{ij}^*$, is increased; see eq. (A2a) in the Appendix. The transition temperature to the topologically ordered low temperature phase is found to be robust against changes in the $a_{ij}^*$ parameter. The strength of the short range, non-electrostatic force does not affect the phase transition precisely because the transition is driven by the electrostatic interaction. This is why $T_C^*$ is unaffected by $a_{ij}^*$, as long as the charge $q^*$ remains constant; see Fig. 1(a). The curves in Fig. 2(a) show the existence of *a single universal curve*, where the critical temperature is the same, regardless the value of $a_{ij}^*$, $T_C^* = q*^2/4r_C^*$, as for the KT transition [1]. Figure 2(b) displays the temperature dependence of the TOP when the thickness of the simulation box along the *z* – axis is increased while keeping the transversal area on the *xy* – plane fixed. The purpose of performing these calculations is to determine to what extent the properties of the topological phase transition are affected as the system becomes more three–dimensional [5]. The results show that the magnitude of the TOP becomes smaller with increasing box thickness. However, the transition temperature remains roughly constant. Extrapolating the trends seen in Fig. 2(b) indicates that the transition disappears when the system transitions from quasi-2d to 3d, in accordance with expectations [45].

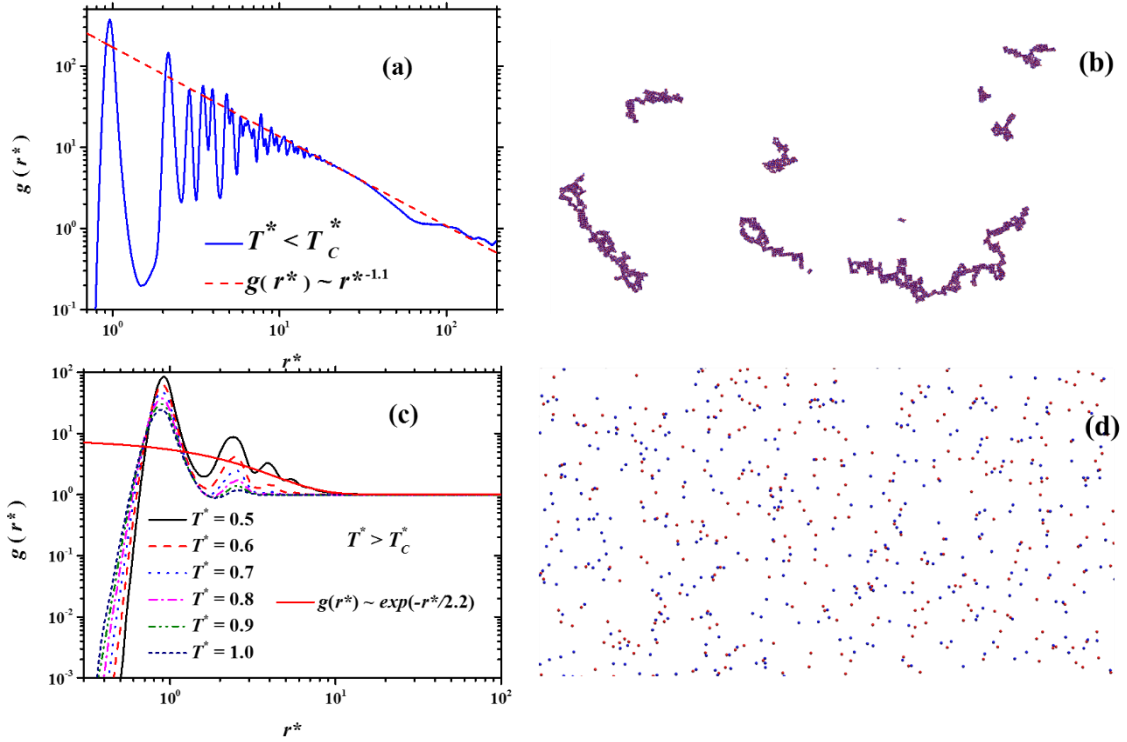

**Figure 3.** (a) The RDF and (b) snapshot of oppositely charged spheres at $T^*/T_c^* = 0.24$. The RDFs shown in (c) are all for $T^* > T_c^*$, with $T_c^* = 0.41$. The snapshot in (d) shows the system at $T^*/T_c^* = 2.45$. The dashed red line in (a) is the fit to the function $g(r^*) \sim r^{*-\eta}$, with $\eta = 1.1$. The solid red line in (c) is the function $g(r^*) \sim e^{-(r^*/2.2)}$. The snapshot in (d) is taken at $T^* > T_c^*$, showing most charges unpaired. Red/blue circles are positive/negative charges. In all cases, $\rho^* = 0.03$, $|q^*| = 7$, $N = 3 \times 10^4$. The snapshots were obtained with VMD [46].

The radial distribution function (RDF), $g(r)$, for particles of opposite charge is shown in Fig. 3, at $T^* < T_C^*$ (Figs 3(a) and (b)), and at $T^* > T_C^*$ (Figs. 3(c) and (d)). At the lowest temperatures, all dipoles are bound in dipole pairs, see Fig. 3(b). The sharp maxima displayed at relatively short distances in the RDF (Fig. 3(a)) correspond to closely bound dipolar pairs [15]. Accordingly, maxima appear at $r^* = 1, \sqrt{5}, \sqrt{8}, ...$, etc., as a consequence of the coordination of opposite charges forming a square lattice on the *xy*–plane, with the first peak due to dipolar pairs, the second to four–charge clusters, and so on. The long–range behavior of the RDF, indicated by the dashed line in Fig. 3(a), shows algebraic decay, as expected [1–4]. To capture this dependence, the area of the simulation box has to be large enough to allow for the formation of dipolar clusters. If the number of charges is small, they all condense into a single square lattice at the lowest temperature. The snapshot in Fig. 3(b) shows all charges condensed into clusters with quasi–long range order at the lowest temperature. At $T^* > T_c^*$, the RDF decays exponentially [1–3], see Fig. 3(c). The charge clusters are dissolved, and most charges move individually, although some remain bound in dipoles; see Fig. 3(d). Comparison of the RDFs below (Fig. 3(a)) and above (Fig. 3(c)) $T_c^*$ vividly sheds light on the nature of this phase transition [1]. Most of the structure of the system when it is condensed, at $T^* < T_c^*$ (Figs. 3(a), (b)) vanishes at $T > T_c^*$ (Figs. 3(c), (d)).

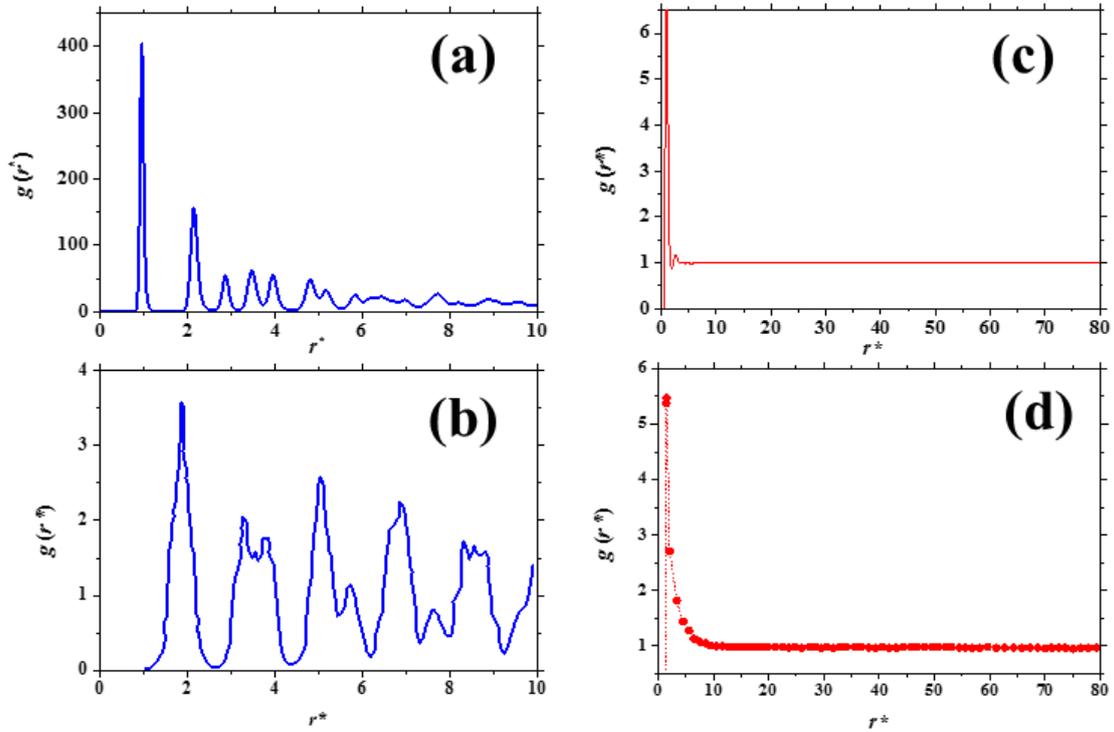

**Figure 4.** Comparison of the radial distribution functions ($g(r^*)$) obtained in this work below (a) and above (c) $T_c^*$, for oppositely charged spheres under quasi – 2d confinement, with those taken from the literature for strictly 2d disks, also below (b) and above (d) $T_C^*$. The data in (a) and (c) are for number density $\rho^* = 0.03$, charge $|q^*| = 7$, and $N = 3 \times 10^4$ at normalized temperature $T^*/T_C^* = 0.24$, and $T^*/T_C^* = 2.44$, respectively. The data in (b) are taken from Clerouin and coworkers [15] for hard charged disks in 2d below $T_C^*$ for $\rho^* = 0.01$ and $T^* = 1/5.7$. The data in (d) are taken from Aupic and Urbic [13] and correspond to a system of hard disks for $\rho^* = 0.01$ and $T^* = 0.6$, above $T_C^*$.

To contrast our predictions with those for strictly 2d systems we compare in Fig. 4 the radial distribution functions obtained in this work between opposite charges below and above the transition temperature, Figs. 4(a) and 4(c), respectively, with those obtained for disks in 2d. Figure 4(b) shows the radial distribution function obtained from molecular dynamics simulations of charged disks interacting through the potential $v_{12}(r) = e^2[(\sigma/r)^\nu + \ln(r/L)]$, where the exponent $\nu = 10$, $\sigma$ is the core radius of the disks, $L$ is and arbitrary length scale, and $e$ is the elementary charge [15]. The positive ions are fixed on a hexagonal cell and the negative ions are free to move. The data shown in Fig. 4(b) were obtained [15] below the transition temperature, when most negative ions are grouped near the positive ions, thereby giving rise to maxima in the radial distribution function, as in Fig. 4(a) for spheres. The periodicity of the peaks in Fig. 4(b) [15] is different from our results, Fig. 4(a), because in this work all charges are free to move whereas the positive charges in the work of Clerouin et al. [15] are fixed on a hexagonal lattice. Figure 4(d) displays the radial distribution function for oppositely charged hard disks obtained from Monte Carlo simulations above the transition temperature [13]. The spatial correlations are seen to decay exponentially for disks, as they do for spheres, see Fig. 4(c), above the transition temperature. There is only a very weak correlation at small distances but that is because the DPD intermolecular interactions of the spheres are not as short-ranged as those of hard spheres; see eq. (A2a) in the Appendix. In conclusion, the spatial correlations for spheres under quasi–2d confinement below (Fig. 4(a)), and above (Fig. 4(c)) $T_C^*$ present qualitatively the same behavior as their strictly 2d counterparts (Figs. 4(b), 4(d)).

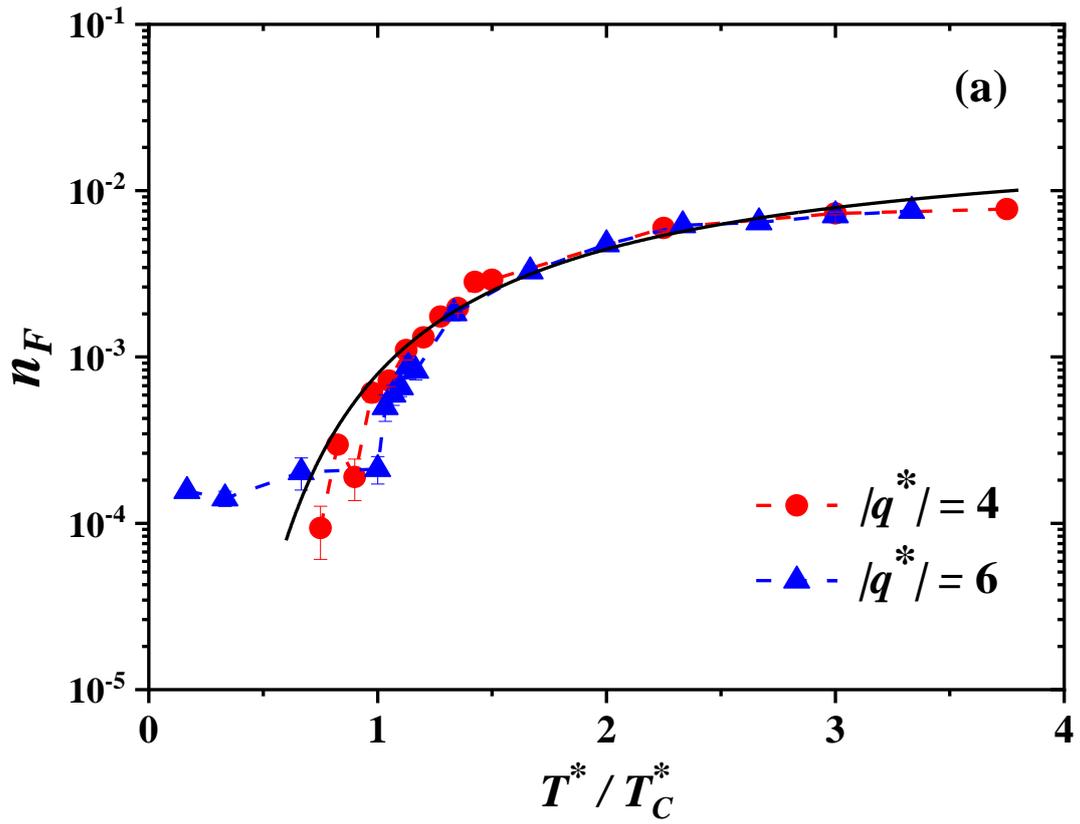

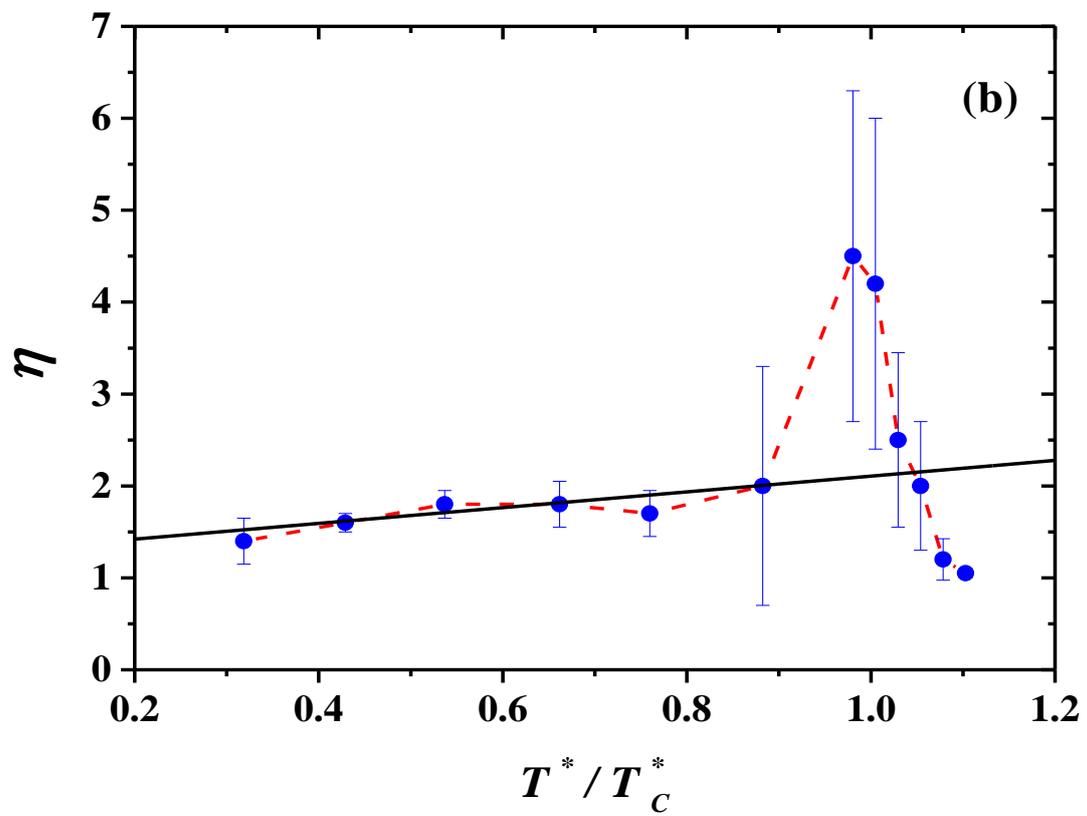

**Figure 5.** (a) Number density of free charged spheres, $n_F$, regardless of their sign, as a function of reduced temperature for spheres with charge $|q^*| = 4$ (red circles) and $|q^*| =$

6 (blue triangles). In both cases there are $N = 200$ charged particles, with total number density $\rho^* = 0.03$. The solid black line represents an approximate analytical solution for finite screening length of the Coulomb interaction between disks taken from [3]. The function is $n_F = a(r_0/\lambda)^{1/2(T/T_C)}$, where $r_0$ is the linear extension of the charge distribution, $\lambda$ is the screening length and $a$ is a parameter [3], chosen here to be $a = 0.025$, with $(r_0/\lambda) = 0.001$. (b) Exponent ($\eta$) of the algebraic decay of the radial distribution function between charges of opposite sign, $g(r_{ij}^*) \sim 1/r_{ij}^{*\eta}$, as a function of the normalized temperature. The solid black line is the best linear fit for the data below $T_C^*$. There are $3 \times 10^4$ particles in the system with $|q^*| = 7$ and $\rho^* = 0.03$.

In Fig. 5(a) we show the evolution of the density of free charges with increasing temperature for two values of the charge. The charges are considered free if their relative distance is $r_{ij}^* \geq 1.4$. This condition was chosen based on the value of the cutoff length of the DPD forces, $r_C^* = 1.0$, see eqs. (A2) in the Appendix. For relative distance between particles larger than $r_C^*$, all the non-bonding, non-electrostatic forces vanish. Thus, at any relative distance $r_{ij}^* > r_C^*$ the charged particles can be considered unbound. By choosing particles to be free if $r_{ij}^* \geq 1.4$, ample range is provided to ensure that under such condition they are not bound. As Fig.5(a) shows, the number of free charges increases rapidly as the temperature grows above $T_C^*$, following the same trend for both values of the charge on each sphere. The solid line in Fig. 5(a) corresponds to the approximate analytical solution, provided by Minnhagen [3] for the Kosterlitz – Thouless transition in charged disks. It is concluded that the quasi 2d condensed phase melts similarly as the strictly 2d phase does. At temperatures below the transition temperature, the spatial correlations between charges of opposite sign decay as a power law, $g(r_{ij}^*) \sim 1/r_{ij}^{*\eta}$, where the exponent $\eta$ depends linearly on temperature, for $T^*$ well below $T_C^*$; see Fig. 5(b). This is in agreement with the expected behavior for the strictly 2d KT transition [1]. At $T_C^*$, $\eta \approx 4.5 \pm 1.8$, which is much larger than the expected value for the KT transition, $\eta = 1/4$ [1]. A more appropriate comparison would be with the value of the exponent extrapolated from the low temperature data (black line in Fig. 5(b)), where the power law dependence of $g(r_{ij}^*)$ is well defined. There, one finds $\eta = 2.1$. Although notably larger than the expected 2d value, this value of $\eta$ is not too different from that found in other, quasi – 2d models [47, 48]. We note that the $1/r$ interaction between charged spheres that we investigate decays more rapidly than the $\ln(1/r)$ interaction, appearing in the BKT theory, between two-dimensional defects (including notably *XY* vortices) or charges. Since the $1/r$ interactions drop more swiftly with particle separation than the conventional logarithmic interactions, the associated correlations that they lead to may drop more precipitously with distance than those in the canonical BKT theory.

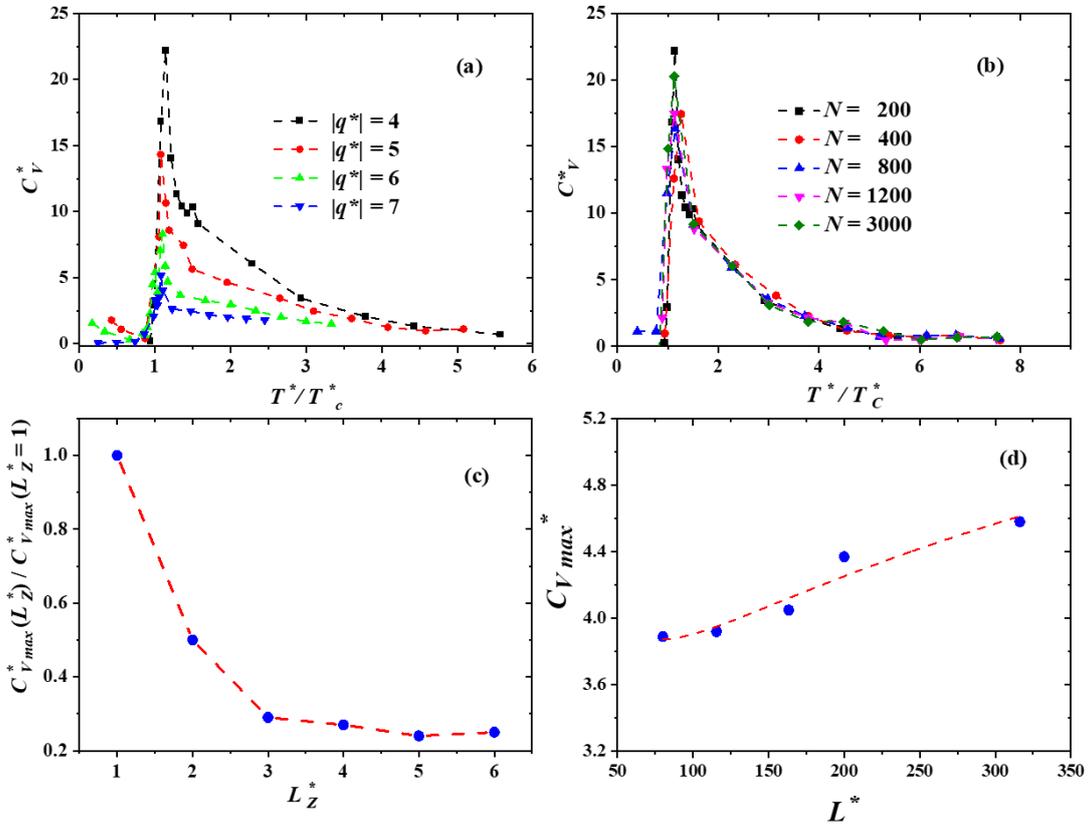

**Figure 6.** (a) Specific heat, $C_V^* = 1/Nq^{*2}\,(\partial U^*/\partial T^*)_V$, as a function of temperature normalized by $T_C^*$, for four values of the charge. The number of charges is $N = 200$. (b) Evolution of the specific heat as a function of temperature, as the number of charged spheres, $N$, is increased. The charge is $|q^*| = 4$. (c) Maximum value of the specific heat as a function of box thickness, normalized by the maximum in $C_V^*$ for the thinnest simulation box, $L_Z^* = 1$. The charge is $|q^*| = 4$, and the transverse area of the box is equal to $L_X^* \times L_Y^* = 80 \times 80 r_C^2$. (d) Dependence of the maxima in the specific heat ($C_{V\,max}^*$) of spheres under quasi–2d confinement as the length of the side of the square area ($L^*$) of the parallelepiped simulation box on the $xy$ – plane is increased. The number density is equal to $\rho^* = 0.03$ and the charge is $|q^*| = 7$. The red dashed line is the best fit to the function $C^*_{V\,max} = a + b\ln(L^*) + c/L^*$ [49]. In all cases the number density is equal to $\rho^* = 0.03$; in (a), (b) and (d) the thickness of the box along the $z$ – direction is fixed at $L_z^* = 1$.

The transition found here can be understood as being driven by the proliferation of defects that melt the crystal [1, 50, 51]. As shown by Toxvaerd [52], the lattice melts as the temperature increases, where the defects become charge quadruplets of disclinations. The role of disclinations is played by dipolar pairs. We calculate the specific heat, $C^*_V = 1/Nq^{*2}(\partial U^*/\partial T^*)_V$, as $|q^*|$ increases, from numerical differentiation of the total internal energy curves, $U^*$. As seen in Fig. 6(a), we find that $C^*_V$ *has a maximum at* $T^* \sim T^*_c$, regardless of the charge, in agreement with previous reports for strictly 2d systems [9, 11–13, 15, 17]. The maximum in $C^*_V$ occurs at a temperature that increases with $|q^*|$, since larger energy is required to unbind dipoles for larger charge and becomes smaller in magnitude as $|q^*|$ increases. The stronger coupling produced by increasing $|q^*|$ makes the transition broader, with more dipolar pairs present at $T^* > T^*_c$. This behavior reproduces that found for $C^*_V$ of charged disks below and above $T^*_c$ [15]. Once $T^*$ is renormalized by the corresponding $T^*_c$ for each charge, the maximum in $C^*_V$ occurs at roughly the same normalized temperature, for all the values of $q^*$; see Fig. 6(a). The influence on the specific heat of increasing the total number of charged spheres in the system, $N$, while keeping the number density and the charge on the particles fixed is presented in Fig. 6(b). Since the charge is fixed, the temperature at which the specific heat reaches its maximum is relatively unaffected by the change in $N$. This is in agreement with the results of Clerouin and coworkers for charged disks [15]. Just as the structure of the system shows that the topological phase transition disappears as it becomes more 3d (see Fig. 2(b)), the thermodynamics of the system confirms this fact. This is more clearly seen in Fig. 6(c), which shows that the maximum in the specific heat (normalized by its value for the thinnest box, with $L_z^* = 1$) is reduced as the box thickness increases. The

finite – size scaling analysis of Fig. 6(d)) shows that the maximum in the specific heat as the temperature crosses the transition temperature for charged spheres under quasi–2d confinement displays a weak dependence on the size of the system. The dashed line in Fig. 6(d) is the best fit to the function $C^*_{V\,max} = a + b\ln(L^*) + c/L^*$ [49], where $a$, $b$ and $c$ are fitting parameters and $L^*$ is the side length of the square transversal area of the simulation box. This fitting function is consistent with the scaling exponent $\alpha = 0$ expected for the 2d Ising model, where the scaling relation is $C^*_V \sim |T^* - T^*_C|^{-\alpha}$ [53]. For the 2d $XY$ model the maximum in the specific heat occurs at a temperature higher than $T^*_C$, signaling the proliferation of defects [1]. In our work the maximum in the specific heat occurs also at temperatures slightly larger than $T^*_C$, with the difference between these temperatures becoming larger as the value of the charge increases.

We also calculate the average current in the system, $I^*_x$, once a weak electric field $\vec{E}^* = 0.02\hat{x}$ is applied along the $x$–axis of the simulation box, for particles with charge $|q^*| = 5$. The current is given by

$$I^*_x = \frac{1}{Nr^*_C}\langle |\sum_{i=1}^{N} q^*_i \vec{v^*}_{ix}|\rangle, \qquad (2)$$

where $\vec{v^*}_{ix}$ is the $x$–component of the velocity of the $i$–th particle with charge $q^*_i$, and $r^*_C = 1$ is the cutoff length of the non–electrostatic forces. The results are shown in Fig. 7, where $U^*$ is included also, for comparison. There is clearly a transition from a low temperature dielectric phase to a conducting phase at $T^* > T^*_C$, as in the strictly 2d KT transition. Consistent with Ohm´s law, the temperature dependence of the current follows that of $U^*$. The inset in Fig. 7 presents $U^*$ vs $I^*_x$, where a linear dependence between them clearly arises. The trends found in both $C^*_V$ and $U^*$ in Fig. 7 are the same as those found for charged disks in strictly 2d systems [9, 14].

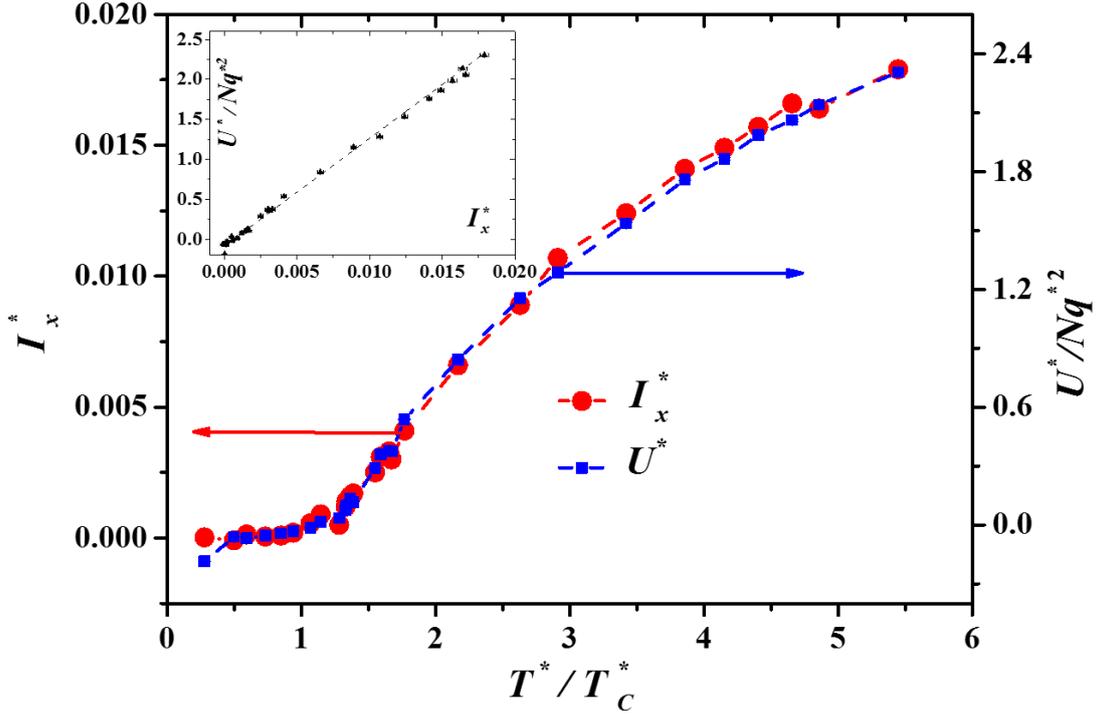

**Figure 7.** Comparison of the internal energy, $U^*/Nq^{*2}$, (blue squares, right axis) with the average current along the direction of the applied electric field, $I_x^*$, (red circles, left axis), when $E^* = 0.02$ is applied along the *x* axis. The dashed lines joining the data are guides for the eye. The inset shows the dependence of $U^*/Nq^{*2}$ on $I_x^*$; the dashed line is the best linear fit. For all curves, $\rho^* = 0.03$, $N = 200$, $|q^*| = 5$ and the error bars are smaller than the symbol sizes.

The phase transition found here may, once again, be understood as being driven by a proliferation of defects in the form of pairs of opposite charges that melt the crystal. At the onset of the melting transition, bound dislocation pairs begin to appear. Raising the temperature leads to the appearance of more dislocations and disclinations, with some remaining at the largest temperature modeled here. This scenario is similar to the KTHNY theory [1, 50, 51] that predicts that the 2d crystal melts due to the unbinding of dislocations and the spread of disclinations. Consequently, the effective interactions between defects responsible for the phase transition in this quasi–2d system may be predominantly logarithmic, as they are in the KTHNY theory, although the electrostatic pair interactions vary as $1/r$.

Lastly, Fig. 8 illustrates the structural evolution of a system of 200 charged spheres on a quasi – 2d simulation box as the temperature is increased, with a series of snapshots. The melting of the ordered structure is found to occur at first by the appearance of defects as pairs of charges (see, for example, the rightmost snapshot in the top row in Fig. 8). As the temperature is raised, the charges remain in groups of four or at least as dipoles mostly, breaking the translational order; see the third and fourth row snapshots in Fig. 8. This transition has features reminiscent of the KTHNY transition in strictly 2d systems [54, 55]. At very high temperatures the dipoles decouple, and the system becomes conducting.

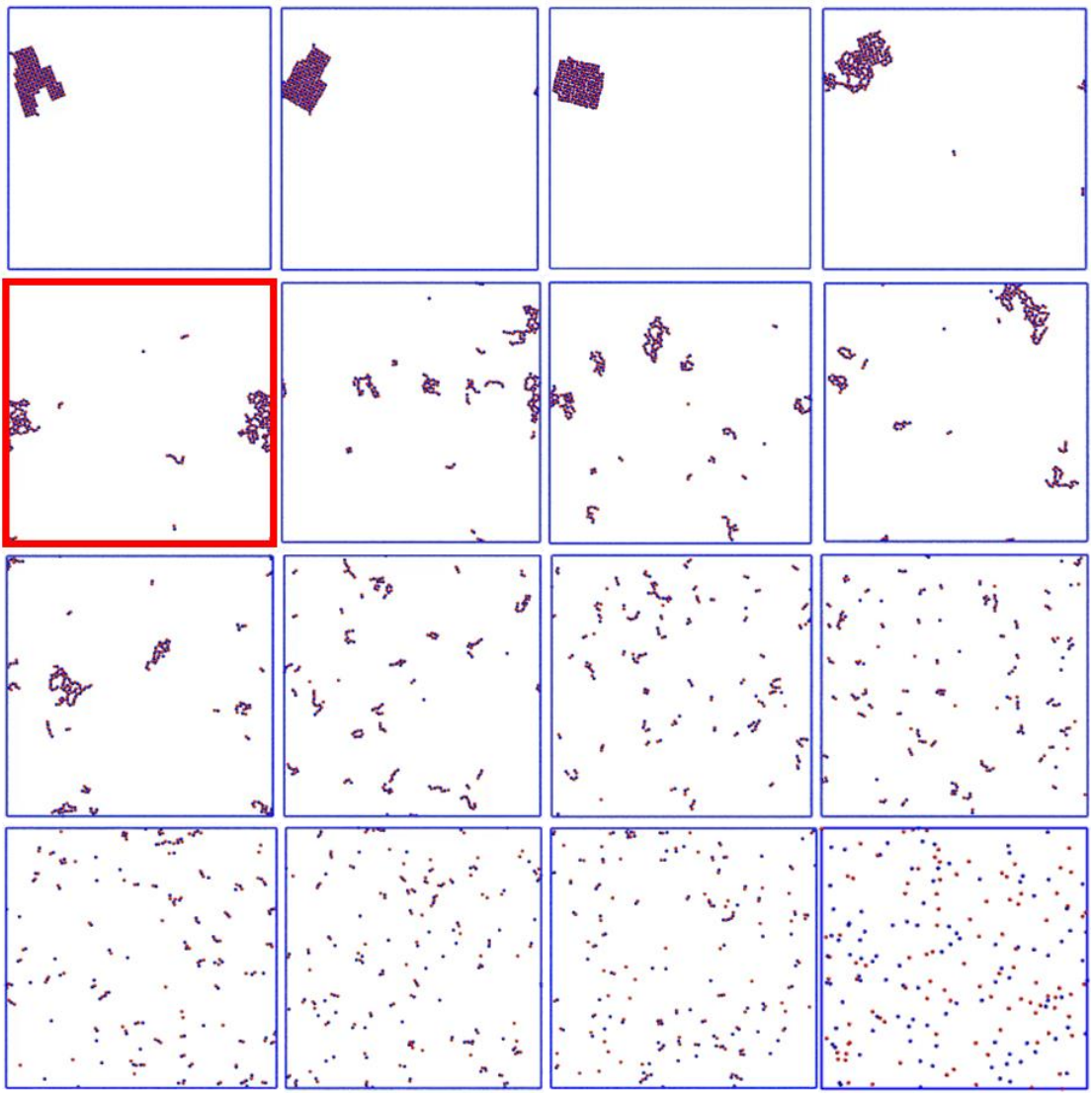

**Figure 8.** Snapshots on the *xy*–plane of the system with charge $|q^*| = 7$ and number density $\rho^* = 0.03$ as a function of temperature, showing how the crystal – like structure

at low temperature melts as the temperature is increased. The normalized temperatures of the first row are, $T^*/T_C^* = 0.24, 0.49, 0.61, 0.98$. For the second row, $T^*/T_C^* = 1.0, 1.02, 1.05, 1.07$. For the third-row snapshots, $T^*/T_C^* = 1.1, 1.22, 1.46, 1.71$. For the last row: $T^*/T_C^* = 1.95, 2.2, 2.44, 122.4$; in all cases, from left to right. The blue line shows the edges of the simulation box, whose volume is $V^* = 80 \times 80 \times 1 r_C^{*3}$. In all cases there is a total of 200 charges; red points are positively charged spheres and blue ones are negatively charged. The system at the critical temperature is emphasized by the frame in red. There are periodic boundary conditions on the *xy*–plane. The snapshots were obtained with VMD [46].

## IV CONCLUSIONS

To summarize, the structural, thermodynamic, and dynamic properties of neutral systems of charged spheres confined to move in a quasi–2d geometry have been obtained at various temperatures, for various charge values. We find that it is possible to locate a topological phase transition in a low-density Coulomb gas that is not strictly confined to 2d but has a finite thickness, and that its properties resemble those of its strictly 2d counterpart. The $T_c^*$ is found to take place at $T_C^* = q^{*2}/4r_C^*$, as predicted for disks [1–8]. The explicit dependence of the Coulomb interaction (ln(*r*) or 1/*r*) is not crucial for the appearance of the topological phase transition but charge neutrality in reduced dimensionality is [11]. This is relevant for comparison with, understanding of and interpretation of recent quasi–2d experiments. Our approach may suggest new experimental tests of our predictions, such as in colloids trapped between membranes, as well as new theoretical approaches to probe this type of topological phase transition.

## V ACKNOWLEDGEMENTS

A.G.G. acknowledges J. D. Hernández Velázquez for enlightening discussions, and the computer resources offered by the Laboratorio de Supercómputo y Visualización en Paralelo (LSVP − Yoltla) of UAM − Iztapalapa, where the simulations were carried out. This project was sponsored by CONACYT through grant number 320197.

## APPENDIX

# AI. THE DPD MODEL

The standard algorithm of molecular dynamics [44] is used for the dissipative particle dynamics (DPD) force model [37, 38]. The net (non – electrostatic) force acting on the $i$ – th particle, $\boldsymbol{F}_i$ is given by:

$$\boldsymbol{F}_i = \sum_{i \neq j} \boldsymbol{F}_{ij}^C + \boldsymbol{F}_{ij}^D + \boldsymbol{F}_{ij}^R \ , \tag{A1}$$

where the DPD interparticle force exerted by particle $i$ on particle $j$ is pairwise additive. The conservative $\boldsymbol{F}_{ij}^C$, dissipative $\boldsymbol{F}_{ij}^D$, and the random forces $\boldsymbol{F}_{ij}^R$ are defined as follows, respectively:

$$\vec{F}_{ij}^C(r_{ij}^*) = a_{ij}^*(1 - r_{ij}^*/r_C^*)\Theta(r_C^* - r_{ij}^*)\hat{r}_{ij} \tag{A2a}$$

$$\vec{F}_{ij}^D(r_{ij}^*) = -\gamma(1 - r_{ij}^*/r_C^*)^2[\hat{r}_{ij} \cdot \vec{v}_{ij}^*]\Theta(r_C^* - r_{ij}^*)\hat{r}_{ij} \tag{A2b}$$

$$\vec{F}_{ij}^R(r_{ij}^*) = \sigma(1 - r_{ij}^*/r_C^*)\Theta(r_C^* - r_{ij}^*)\xi_{ij}\hat{r}_{ij}. \tag{A2c}$$

The length $r_C^* = 1$ is a cutoff radius and $\Theta(x)$ is the Heaviside step function. Here, $\vec{v}_{ij}^* = \vec{v}_i^* - \vec{v}_j^*$ is the relative velocity between particles $i$ and $j$. Due to the fluctuation–dissipation theorem, $\sigma^2/2\gamma = k_B T$, where $k_B$ is Boltzmann's constant and $T$ is the absolute temperature [38]. The amplitudes $\xi_{ij}$ are randomly drawn from a uniform distribution between 0 and 1. They become zero for $r_{ij}^* > r_C^*$, where $r_C^*$ is a cutoff radius. This cutoff distance is selected as the reduced unit of length $r_C^* = 1$ and is the intrinsic length scale of the DPD model. The masses of all particles are equal and chosen as $m_i^* = 1$, in reduced units. Asterisked quantities are reported in reduced units, except where explicitly stated otherwise. Figure 9 illustrates the dependence of the non–bonding, non-electrostatic conservative DPD force ($F_{ij}^C$) on the relative distance $r_{ij}^*$ between the centers of mass of the particles. For distances larger that the cutoff distance, $r_C^*$, this force vanishes, and the particles behave as if they were in an ideal gas. For relative distances in the range $0 \leq r_{ij}^* \leq r_C^*$ the force grows linearly as the distance is reduced.

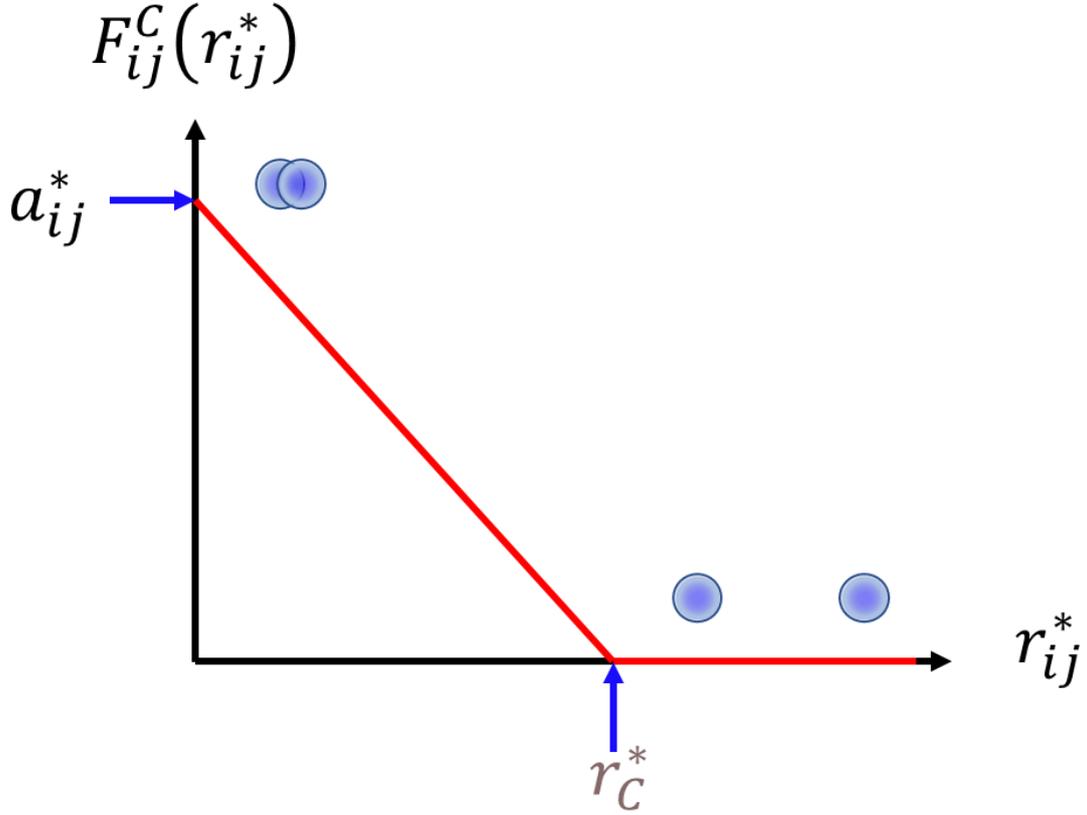

**Figure 9.** Non–bonding, pairwise conservative DPD force ($F_{ij}^C$, red line) as a function of the relative distance between the centers of mass of the particles ($r_{ij}^*$); see eq. (A2a). The maximum intensity of this repulsive force if given by the constant $a_{ij}^*$, and $r_C^*$ is the cutoff radius. The blue circles are cartoons representing particles overlapping when their relative distance is smaller than $r_C^*$, when $F_{ij}^C$ is large but finite. For relative distances larger than $r_C^*$, $F_{ij}^C = 0$.

The constant that defines the maximum amplitude of the random force, $\sigma$, is in all cases fixed at $\sigma = 3$, as is customary [38]. Once the temperature is chosen, the amplitude of the dissipative force, $\gamma$, is fixed, since $\gamma = \sigma^2/2k_BT$ [38]. The constant that defines the maximum strength of the non – electrostatic conservative DPD force, $a_{ij}^*$, is chosen as $a_{ij}^* = 78.3$ for all particles, regardless of their charge, except when stated otherwise. This value is obtained from the equation [39]:

$$a_{ii}^* = \frac{[16N_m - 1]}{0.6} k_B T^* \quad , \tag{A3}$$

where $N_m$ is the so called "coarse graining degree", i.e., the volume of the DPD particles in terms of the volume of a water molecule; we choose $N_m = 3$ [39]. In reduced units, the thermal energy scale is $k_B T^* = 1$. Since this force is linearly decaying with interparticle separation, the particles are soft and could in principle overlap, see Fig. 9. However, by choosing $a_{ij}^* = 78.3$ one finds that particle overlap is always avoided [40].

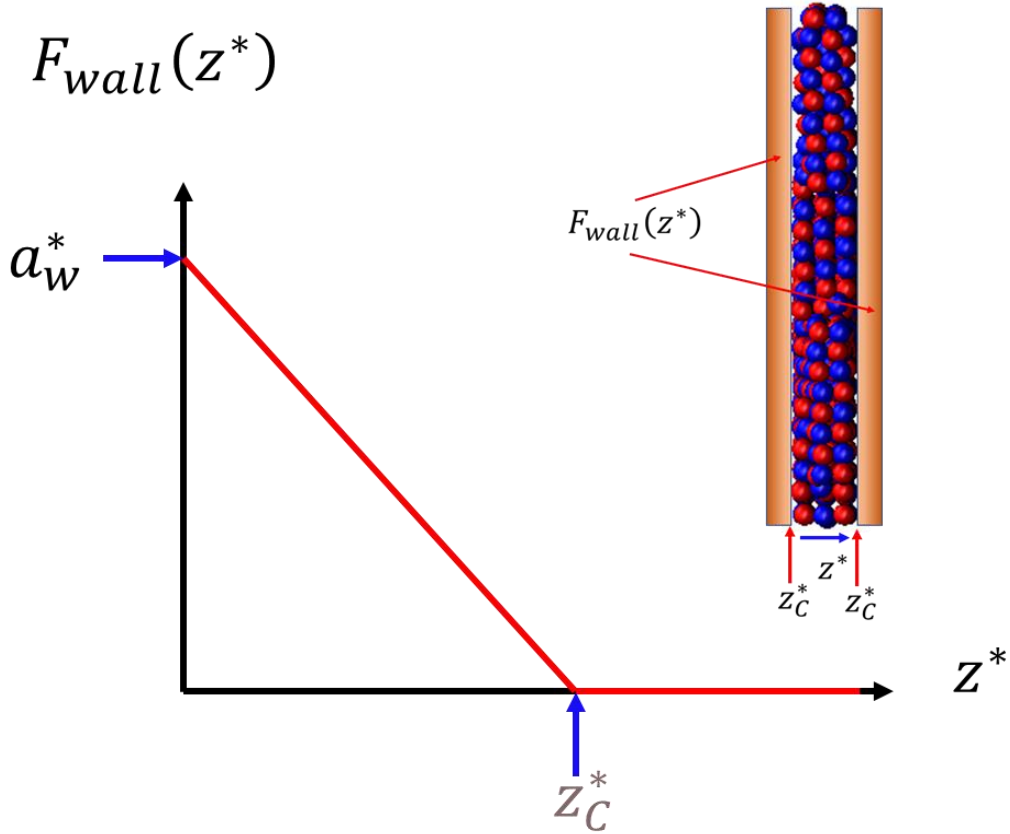

**Figure 10.** Effective wall force applied perpendicularly to the $xy$ – plane at both ends of the simulation box along the $z$ – direction. The wall force is linearly – decreasing with the $z$ – component of the particles' positions. The maximum strength of the wall force is given by the constant $a_w^*$; the force vanishes for $z^* > z_C^*$. The inset is a cartoon showing the walls (brown rectangles), defined by eq. (A4), confining the charged particles in the box. The snapshot of the charges flanked by the walls is taken from an actual simulation, with red circles representing positive charges and blue ones representing negative charges.

To keep the charged spheres moving on a quasi two-dimensional slit on the $xy$ – plane, a force along the $z$ – axis is applied at both ends of the simulation cell on the z – axis. It is defined as [41]

$$\vec{F}_{wall}(z^*) = a_w^*(1 - z^*/z_C^*)\hat{z}, \qquad (A4)$$

where the amplitude of the wall force is chosen in all cases as $a_w^* = 150.0$ in reduced units. This value of the $a_w^*$ constant has been shown to be strong enough so that the particles cannot penetrate the walls [41]. This wall force becomes identically equal to zero for distances along the z–axis larger than $z_c^* = 1$, and it is added to the other non–bonding, non–electrostatic forces in eq. (A1); see Fig. 10. The wall force defined by eq. (A4) confines effectively the charged particles along the z – direction.

## AII. The Coulomb Interaction for Charge Distributions

It was Groot who first proposed to use charge distributions to deal with the softness of DPD particles, thus avoiding singularities in the Coulomb interaction arising from complete overlap between particles [42]. Here we use also charge distributions instead of point charges but follow the model of González–Melchor et al. [43] instead, in which the electrostatics of the DPD particles is solved with the Ewald sums technique [44]. The charge distributions are defined as a Slater–type charge density function, given by

$$\rho_{q^*}^*(r) = \frac{q^*}{\pi \lambda^{*3}} e^{-2r^*/\lambda^*}, \qquad (A5)$$

where $\lambda^*$ is the decay length of the charge. When the charge density in eq. (5) is integrated over space one finds that the total charge in the particle is $q^*$. In general, the forces between charge distributions cannot be calculated analytically. However, for the model given by eq. (A5) there is an accurate approximate expression that has been successfully tested in various applications [56, 57]. The approximate Coulomb interaction between the charge distributions defined by eq. (A5) is [43]:

$$U(r^*) = \frac{\Gamma}{4\pi} \left(\frac{q_i^* q_j^*}{r^*}\right) [1 - (1 + \beta r^*) e^{-2\beta r^*}], \qquad (A6)$$

while the force derived from it is given by

$$\vec{F}_e(r^*) = \frac{\Gamma}{4\pi}\left(\frac{q_i^* q_j^*}{r^{*2}}\right)\left[1 - (1 + 2\beta r^*\{1 + \beta r^*\})e^{-2\beta r^*}\right]\hat{r} \qquad (A7)$$

where $\Gamma = \frac{e^2}{k_B T \epsilon_0 \epsilon_r r_C^*}$ and $\beta = \frac{r_C^*}{\lambda^*}$. The constants $\epsilon_0$ and $\epsilon_r$ are the dielectric constants of vacuum and water at room temperature, respectively. As shown in Fig. 11, both the potential in eq. (A6) and the force derived from it, eq. (A7), are well behaved when the interparticle distance goes to zero.

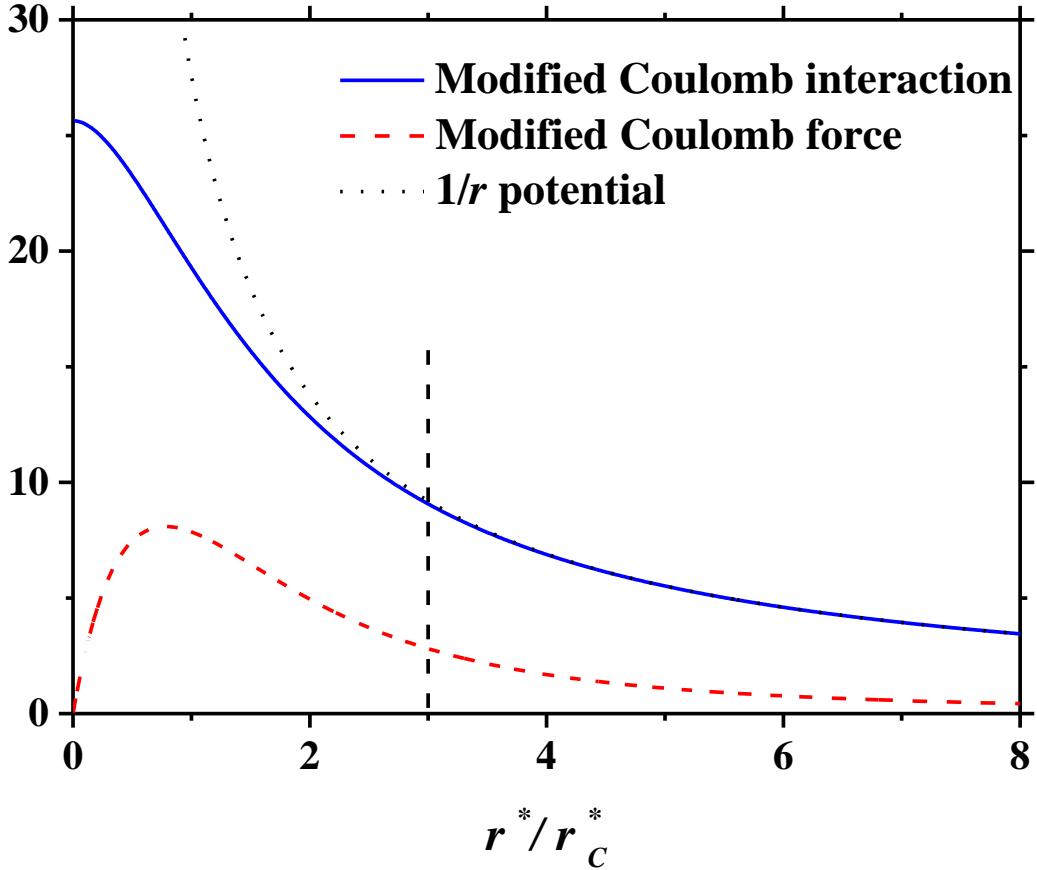

**Figure 11.** Coulomb interaction between charge distributions (solid blue line; see eq. (A6)) and the electrostatic force derived from it (dashed red line); see eq. (A7). The dotted black line represents the bare electrostatic interaction $U(r) \sim 1/r$. The vertical dashed line represents the distance where the real part of the Ewald sums is truncated. For these plots, the value of $|q^*| = 5$ was used. After González – Melchor et al. [43].

For distances smaller that the distance at which the modified Coulomb interaction between charge distributions is equal to the bare Coulomb interaction between point charges ($r^*/r_C^* = 3$, in Fig. 11), the Ewald sums are performed in real space. For distances larger than that, the sums are performed in Fourier space, following the usual procedure [44]. For additional details about the application of Ewald sums to charge distributions in bulk DPD systems, the reader is referred to the work of González–Melchor et al. [43].

When Ewald sums are applied to confined systems, as is our case, additional care must be taken because the Fourier transforms involved cannot be performed straightforwardly due to the lack of three – dimensional periodicity [58, 59]. This problem can be overcome by applying an additional force along the z – axis to all the charged particles:

$$\vec{F_i}(z) = -\frac{\Gamma}{V^*} q_i^* M_z \hat{z},  \quad (A8)$$

where

$$M_z = \sum_{i=1}^{N} q_i^* z_i^*.  \quad (A9)$$

In eq. (A8) $V^*$ is the volume of the simulation cell and $M_z$ is the total dipole moment, see eq. (A9), which must be removed out of the cell for each particle. By applying this additional force to each particle along the z – direction it can be shown [59] that the three – dimensional version of the Ewald sums can be successfully applied to confined DPD particles with charge distributions. Full details can be found in the work of Alarcón et al. [59].

## AIII. Simulation Details

All the simulations reported here are performed in reduced units and under canonical ensemble conditions. The time step chosen to integrate the equation of motion is $\delta t^* = 0.01$; the volume of the simulation box is $L_x^* \times L_y^* \times L_z^* = 80 \times 80 \times 1 r_C^{*3}$, except where indicated otherwise; $r_C^* = 1$ throughout this work. To express distances $r$ in units of length one has to multiply $r = r^* r_C$, with $r_C = 6.46$ Å [60]. The total number of particles in the systems modeled is in the range from $2\times 10^2$ up to $3\times 10^4$ DPD spheres in the simulation box. Periodic boundary conditions are applied along the $x$– and $y$– directions of the simulation box but not along the $z$–direction because the system is confined in that direction. The simulations are carried out using a code developed by the group of one of us (AGG), which has been benchmarked and thoroughly tested [60]. The real part of the Ewald sums is cut off at $r_E^* = 3.0 r_C^*$, and $\alpha = 0.15$ Å$^{-1}$. The latter is the factor that determines the contribution in real space of the Ewald sums [44]. The value for $\lambda^*$ in eq. (A5) is chosen in all cases as $\lambda^* = 6.95$ Å. For the part of the Coulomb interaction calculated in Fourier space, a maximum reciprocal vector $\vec{k}^{max} = (5,5,5)$ is chosen. The constants $\beta$ and $\Gamma$ in eq. (A6) are chosen as $\beta = 0.929$ and $\Gamma = 13.87$, following González–Melchor et al. [43] and Groot [43], respectively. To confine the charged spheres with the surfaces, given by eq. (A4), no periodicity at the boundaries along the $z$-direction is applied, therefore the charges are not replicated along the $z$-axis.

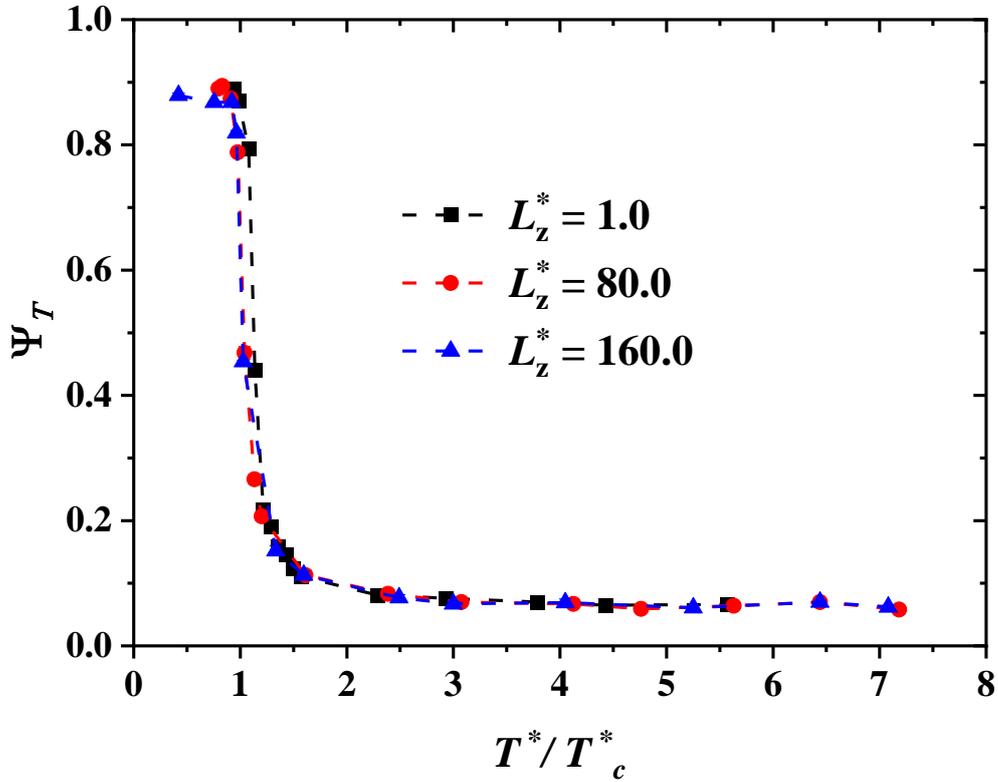

**Figure 12.** The translational order parameter $\Psi_T$ as a function of temperature, normalized by the critical temperature, for a system of 200 charged spheres in boxes of three thickness along the *z*-direction. The data labelled "$L_z^* = 1.0$" (black squares) are taken from Fig. 1(a). In all cases, the distance between the effective walls that confine the charged spheres along the *z*-direction is $\Delta z^* = 1.0$; the rest is vacuum.

The lack of periodicity along the *z*-direction and the impenetrable walls avoid the replication of the charges along the *z*-axis. To further prove this issue, additional simulations were carried out for new box sizes, namely $L_x^* \times L_y^* \times L_z^* = 80 \times 80 \times 80$, and $80 \times 80 \times 160$. The walls were placed at the center of the new boxes along the *z*-axis, with the separation between the walls kept always equal to $\Delta z^* = 1.0$, to maintain the system under a quasi-2d geometry. This setup creates a large vacuum space in either side of the confined charges, along the *z*-direction. A system of 200 charged spheres, each with charge $|q^*| = 4$, was run in each box and the translational order parameter (TOP) was calculated, as a function of temperature, for $2 \times 10^6$ time steps. The results, presented in Fig. 12, show that there is virtually no change in the TOP using boxes with and without explicit vacuum regions along the *z*-direction, consistent with the lack of periodic boundary conditions along the *z*-axis. Following Yeh and Berkowitz [58], the three-dimensional version of the Ewald sums is used, with the dipolar correction for confined geometry, see eq. (A9), properly adapted by Alarcón et al. [59] to DPD with charge distributions. Up to 20 *k* vectors were used to test whether the translational order parameter or the electrostatic interaction depended on them, finding that the results are essentially independent of the number of *k* vectors. Thus, to improve the computational speed we used only 5 *k* vectors in all simulations reported here. The simulations are run for at least $5 \times 10^6$ time steps and up to $4 \times 10^7$ time steps, with the first half used to reach equilibrium and the second half used for the production phase. Because there are equal number of positively and negatively charged particles, the net charge in the system is always zero.

## AIV. ADDITIONAL RESULTS AND DISCUSSION

Figure 13 shows the dependence of the TOP on temperature for increasing values of the number density of charged spheres, for $|q^*| = 5$. The TOP reaches a maximum close to $T_c^*$, signaling an ordering of the charged spheres, decaying sharply at $T > T_c^*$, as expected for a phase transition. It also exhibits a monotonic decrease with increasing number density, yet $T_c^*$ does not change, being determined by the charge, which is constant in Fig. 13. This is the case because the densities are low, which makes the dielectric constant $\epsilon = 1$ and $T_c^*$ becomes density-independent [16].

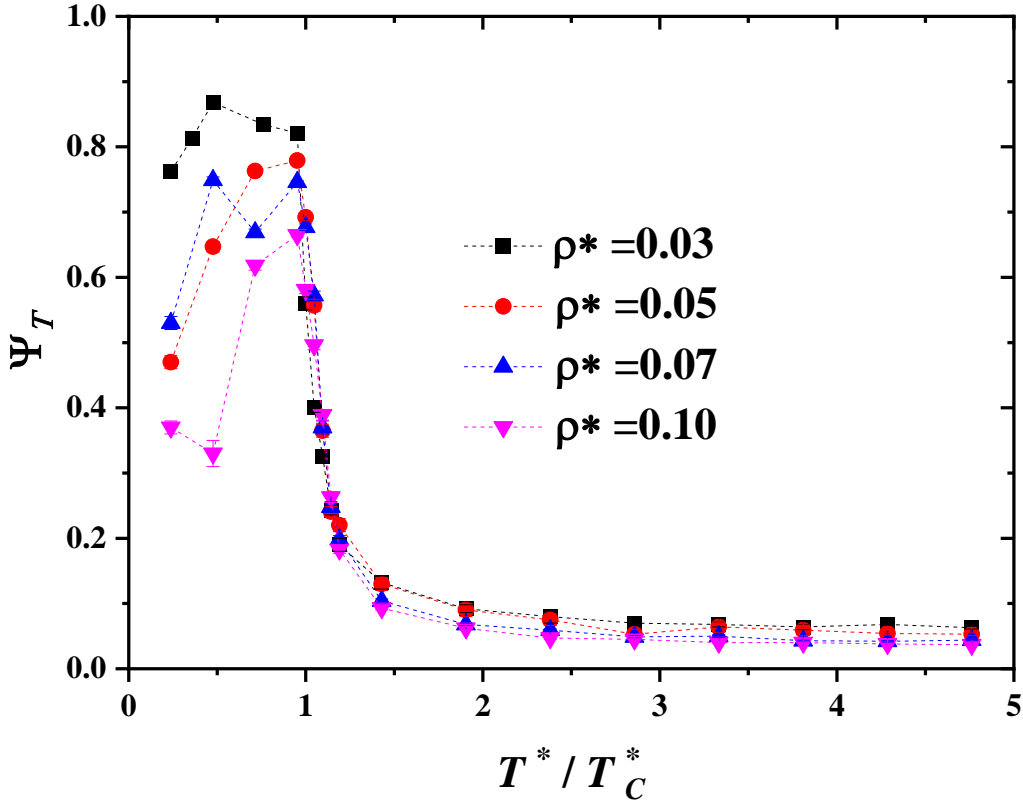

**Figure 13.** Translational order parameter TOP, see eq. (1), *vs* temperature, normalized by $T_c^*$, for four values of the number density of charged particles. In all cases, $N = 200$ and the thickness of the simulation box is $L_Z^* = 1$; error bars are smaller than the symbol's size. The dashed lines are guides for the eye.

The total internal energy, $U^*$, is calculated as well at each temperature modeled and from it the specific heat is obtained as usual, $C_V^* = (1/Nq^{*2}) \partial U^*/\partial T^*$. The derivative is evaluated numerically via finite differences of the internal energy. The results of varying the box thickness are shown in Fig. 13. The specific heat is found to decrease as the box thickness is increased, reaching a plateau for $L_Z^* \geq 3$, which can be considered as the "critical thickness", i.e., when the system ceases to be quasi–two dimensional. As expected, both structural (Fig. 13) and thermodynamic (Fig. 14) properties show that the phase transition tends to disappear as the system becomes more 3d.

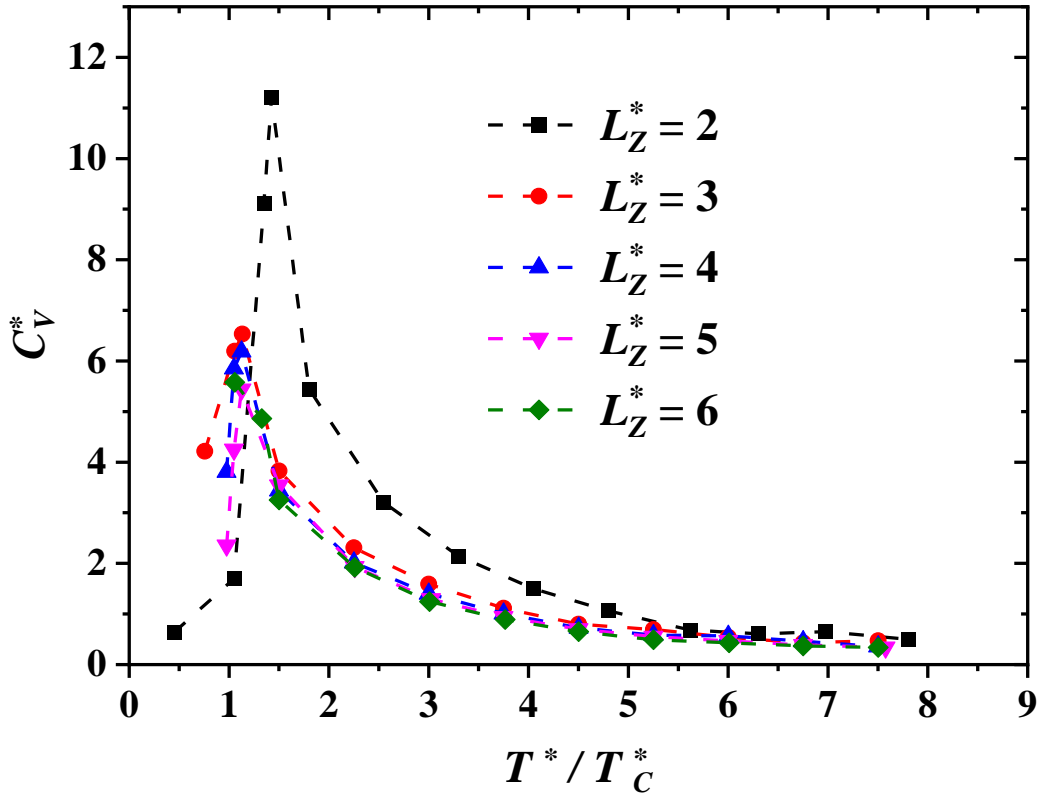

**Figure 14.** Influence of increasing the thickness of the simulation box along the $z$ – axis, $L_Z^*$, on the specific heat. Evolution of the maximum value of the specific heat as a function of box thickness, normalized by the maximum in $C_V^*$ for the thinnest simulation box, $L_Z^* = 1$. In all cases, $|q^*| = 4$, and $\rho^* = 0.03$, and the transverse area of the box is equal to $L_X^* \times L_Y^* = 80 \times 80 r_C^2$. Dashed lines are only guides for the eye.

In Fig. 15 we compare qualitatively our predictions for the specific heat with those obtained by other workers for 2d disks, as we did for the spatial correlations below and above $T_C^*$ (see Fig. 4). In Fig. 15(a) one can find our predictions for the specific heat as a function of temperature for three values of the magnitude of the electric charge on the spheres. Figure 15(b) shows the specific heat data obtained for disks where the positively charged ones are fixed on a hexagonal lattice [15], for different densities. The lines in both panels of Fig. 15 were obtained from numerical differentiation of the internal energy data. The three curves in Fig. 15(b) correspond to different isochores, where decreasing $\sigma$ is equivalent to increasing the charge $q^*$ in Fig. 15(a). Once again it is found that our predictions for charged spheres in quasi–2d conditions reproduce qualitatively the physics of charged disks.

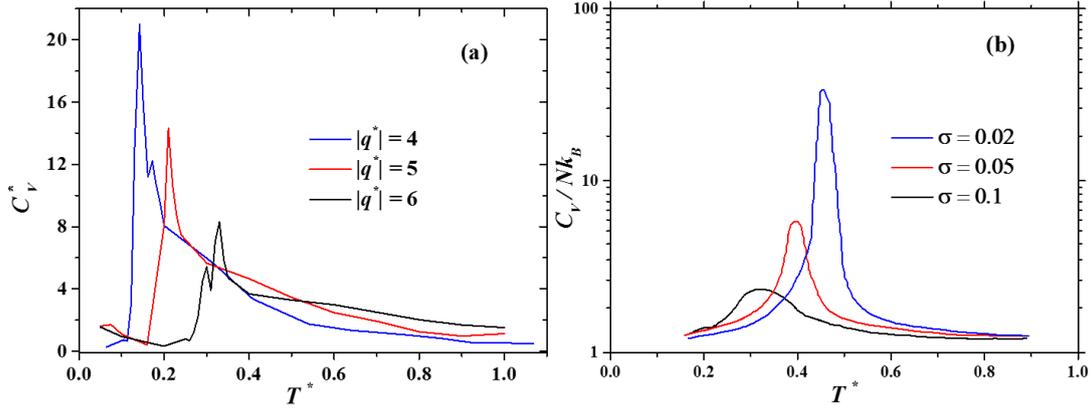

**Figure 15.** Specific heat as a function of temperature for: (a) charged spheres, and (b) charged disks [15]. The data in (a) are reported for three values of the charges on the spheres while keeping the concentration constant at $\rho^* = 0.03$, with the volume of the simulation box given by $L_X^* \times L_Y^* \times L_Z^* = 80 \times 80 \times 1 r_C^3$. The curves in Fig. (b) are taken from Clerouin et al. [15] for strictly – 2d disks. See text for details.

When a weak external electric field is applied to the system of charged spheres confined to move on a quasi – $2d$ environment a phase transition can still be found, as demonstrated by the value of the TOP; see the blue squares in Fig. 16. The temperature at which the transition occurs is raised with respect to the case when the system is in equilibrium (when no external field is applied; see Fig. 1(a)). The system undergoes a transition from a low temperature dielectric phase to a high temperature, electrically conducting phase, as in the strictly 2d case [8].

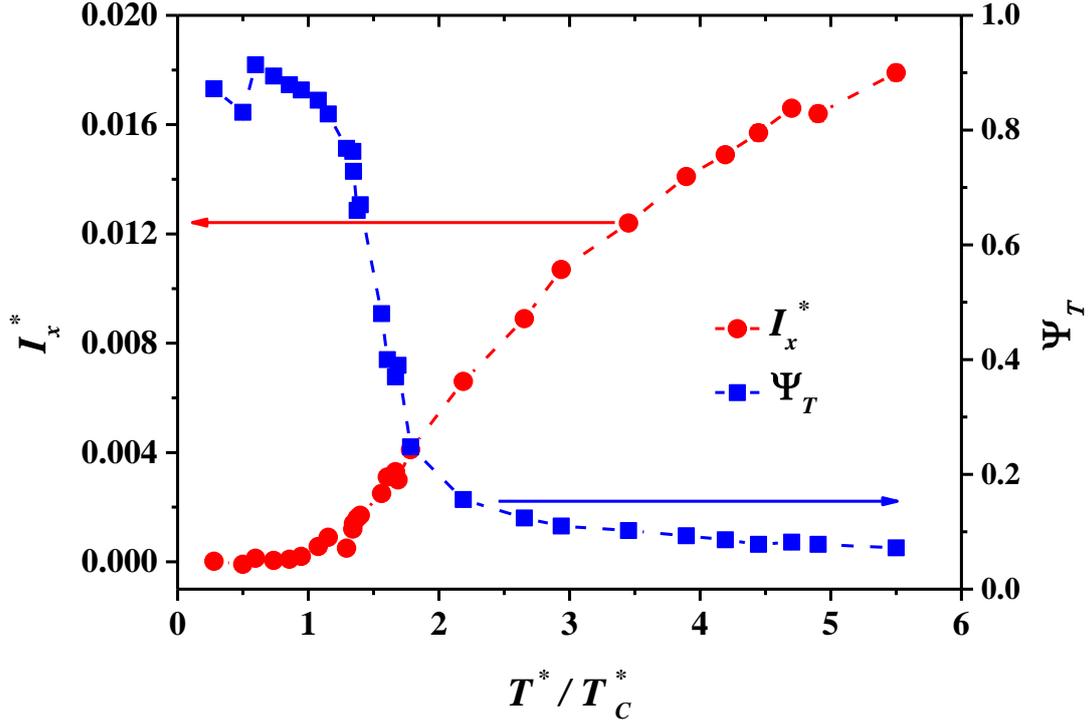

**Figure 16.** Comparison of the translational order parameter (blue squares, right axis) with the electrical current along the $x$ – axis (red circles, left axis) as the temperature is raised. The data correspond to a system of particles with charge $|q^*| = 5$, number density $\rho^* = 0.03$ and simulation box volume $L_x^* \times L_y^* \times L_z^* = 80 \times 80 \times 1 r_C^{*3}$. The electric field applied along the $x$ – direction is $E_x^* = 0.02$. The electric current data are taken from Fig. 7.

The application of a weak electric field yields no current in the low temperature phase, where the order parameter is close to one, while a conducting phase is obtained in the high temperature phase, with vanishing order parameter.